# Analysis on the Integrability of the Rayleigh-Plesset Equation with Painlevé Test and Lie Symmetry Groups

Lang Xia

Email: langxia.org@gmail.com

**Abstract:** The Rayleigh-Plesset equation (RPE) is a nonlinear ordinary differential equation (ODE) of second order that governs the dynamics of a spherical bubble and plays an essential role in interpreting many real-world phenomena involving the presence of bubbles in engineering and medical fields. In the present paper, we present a relatively comprehensive analysis of the analytical solutions of the RPE. The integrability of the Rayleigh-Plesset equation is investigated and discussed using the Painlevé test. Lie symmetry groups are employed subsequentially to obtain several exact solutions to the simplified Rayleigh-Plesset equations.

**Introduction**

The Rayleigh-Plesset equation (hereafter RPE) and its variants frequently appear in the engineering fields with applications to multiphase flows and acoustics (Plesset and Prosperetti 1977, Lauterborn and Kurz 2010). It is the equation that governs the dynamics of microbubbles, many of which are widely used as ultrasound contrast agents. This second-order nonlinear ODE hence is of particular importance to the biomedical imaging (Sarkar, Shi et al. 2005, De Jong, Emmer et al. 2009). Although voluminous numerical investigations of the RPE have been done in the engineering fields (Lauterborn and Kurz 2010), some unexplained phenomena, such as emissions of subharmonics (Faez, Emmer et al. 2013) and nonlinear responses of gas bubbles at linear dynamical range (Xia, Paul et al. 2014, Xia, Porter et al. 2015), may still require analytical solutions for fully understanding its dynamical behaviours. Analytical solutions to the RPE without the surface tension term (named Rayleigh equation) were obtained by using hypergeometry functions (Kudryashov and Sinelshchikov 2015). The closed-form solution to a simplified RPE was also found through Weierstrass elliptic functions (Mancas and Rosu 2016). In both of these studies, the effect of the viscous term was neglected.

Due to the symmetries of solvable equations sensitive to the coefficients in differential equations, finding systematic methods that guarantee the solvability of arbitrary differential equations is not easy (Bryant, Griffiths et al. 1995, Schwarz 2000, Oliveri

2010, Tiwari, Pandey et al. 2013). Additionally, solvable equations may not admit Lie symmetry groups (Muriel and Romero 2011, Morando 2015). These may explain that no analytical solution to the RPE (in this paper Eq.(4)) has been found in the literature by far.

In this paper, we study the integrability and solvability of the RPE with the method of Painlevé test and Lie symmetry analysis (Lie group method), and thus focus on the second-order ordinary differential equation in a geometric perspective, which is in the form of

$$F(t, R, \dot{R}, \ddot{R}) = 0, \quad \dot{R} := \frac{\partial}{\partial t} R, \quad \ddot{R} := \frac{\partial^2}{\partial t^2} R \tag{1}$$

More specifically, we study the Rayleigh-Plesset equation (RPE) for a free gaseous bubble in an incompressible liquid, which may be written as (Franc and Michel 2004)

$$\ddot{R} + \frac{3}{2R} \dot{R}^2 + \frac{4\mu}{\rho R} \dot{R} + \frac{2\gamma}{\rho R^2} - \frac{P_{g0} R_0^{3k}}{\rho R^{3k+1}} + \frac{p(t)}{\rho R} = 0 \tag{2}$$

where $R$ is the instantaneous radius of the bubble, $P_{g0}$ the pressure inside the gas bubble, $R_0$ the initial radius, $k$ the polytropic constant of the inside gas, $\rho$ the density of the surrounding liquid, $\mu$ the viscosity of the liquid, $\gamma$ the surface tension of the liquid, and $p(t)$ the excitation pressure. To focus on the mathematical aspect, we non-dimensionalize the above equation by assuming $R_0$, $\omega$ and $p_0$ to be characteristic scales of bubble radius $R$ ($R \neq 0$), time $t$, and pressure $p$, that is

$$R^* = \frac{R}{R_0}, \quad t^* = t\omega, \quad p^* = \frac{p}{p_0} \tag{3}$$

Thus we have the following non-dimensionalized RPE (the star symbol is omitted)

$$\ddot{R} + \frac{3}{2R} \dot{R}^2 + \frac{1}{\text{Re}\, R^2} \dot{R} + \frac{\text{We}}{R^2} - \frac{p_n}{R^{3k+1}} + \frac{\text{Th}\, p(t)}{R} = 0 \tag{4}$$

where

$$\text{Re} = \frac{\rho R_0^2 \omega}{4\mu}, \quad \text{We} = \frac{2\gamma}{\rho R_0^3 \omega^2}, \quad \text{Th} = \frac{p_0}{\rho R_0^2 \omega^2}, \quad p_n = \frac{p_{g0} R_0^{3k}}{\rho R_0^2 \omega^2} \tag{5}$$

here Re is the Reynolds number, We the Webber number, Th the Thoma cavitation number and $p_n$ the pressure number.

**Painlevé test**

More than one hundred years ago, the French mathematician Paul Painlevé and his colleagues discovered that any second order rational ordinary differential equation without movable branching points could be transformed into the following six Painlevé equations (Noumi 2004)

$$P_1: \quad \ddot{R} = 6\dot{R}^2 + t$$
$$P_2: \quad \ddot{R} = 2R^3 + tR + a$$
$$P_3: \quad \ddot{R} = \frac{1}{R}\dot{R}^2 - \frac{1}{t}\dot{R} + \frac{1}{t}\left(aR^2 + b\right) + cR^3 + \frac{d}{R}$$
$$P_4: \quad \ddot{R} = \frac{1}{2R}\dot{R}^2 + \frac{3}{2}R^3 + 4tR^2 + 2\left(t^2 - a\right)R + \frac{b}{R}$$
$$P_5: \quad \ddot{R} = \left(\frac{1}{2R} + \frac{1}{R-1}\right)\dot{R}^2 - \frac{1}{t}\dot{R} + \frac{(R-1)^2}{t^2}\left(aR + \frac{b}{R}\right) + \frac{c}{t}R + d\frac{R(R+1)}{R}$$
$$P_6: \quad \ddot{R} = \frac{1}{2}\left(\frac{1}{R} + \frac{1}{R-1} + \frac{1}{R-t}\right)\dot{R}^2 - \left(\frac{1}{t} + \frac{1}{t-1} + \frac{1}{R-t}\right)\dot{R}$$
$$+ \frac{R(R-1)(R-t)}{t^2(t-1)^2}\left(a + b\frac{t}{R^2} + c\frac{t-1}{(R-1)^2} + d\frac{t(t+1)}{(R-t)^2}\right)$$

The solutions to the above equations are called *Painlevé transcendents*. We can discuss the integrability of the RPE by checking if there exist movable singularities. We may write the Eq.(4) again in the form of

$$\ddot{R} = -\frac{3}{2R}\dot{R}^2 - \frac{1}{\operatorname{Re}R^2}\dot{R} - \frac{\mathrm{We}}{R^2} + \frac{p_n}{R^{3k+1}} - \frac{\mathrm{Th}p(t)}{R} \tag{6}$$

or

$$F\left(t, R, \dot{R}, \ddot{R}\right) = \ddot{R} - f\left(t, R, \dot{R}\right) = 0 \tag{7}$$

where

$$f = \frac{p_n}{R^{3k+1}} - \frac{3}{2R}\dot{R}^2 - \frac{1}{\text{Re}\,R^2}\dot{R} - \frac{\text{We}}{R^2} - \frac{\text{Th}p(t)}{R} \tag{8}$$

According to the principle of Painlevé test, Eq.(7) is globally integrable if the function $f$ is rational in $\dot{R}$, algebraic in $R$ and analytic in $t$. In Eq.(8) $f$ is not algebraic with respect to $R$ and thus the equation is not globally integrable. Furthermore, an ordinary differential equation is said to have *Painlevé property* if all the movable singularities of all its solutions are poles (Baldwin and Hereman 2006). A singularity is movable if it depends on the constants of integration of the ODE. We can use Laurent series solutions of the differential equation to check if the RPE has Painlevé property. The Laurent series is written in the form of

$$R(t) = \tau^{-\alpha} \sum_{i=0}^{\infty} a_i \tau^i, \quad \alpha \in \mathbb{Z}^+ \tag{9}$$

where $a_0(t) \neq 0$ and $\tau = t - t_0$. The leading order terms (with the lowest exponent of $\tau$) will dominate the behaviors of the solution $R(t)$. By the analysis of the coefficient of the leading term, we can detect if any movable singularities exist. This can be done by checking whether the equation is solvable with respect to the leading term of Eq.(9). Therefore, substituting the following relation

$$R(t) = a_0 \tau^{-\alpha} \tag{10}$$

into the RPE Eq.(6), we obtain the following relation for the leading power as

$$(5\alpha + 2)a_0 \alpha = 0 \tag{11}$$

The solution to the above equation is $\alpha = -2/5$, which is impossible since it has to be an integer. As a result, the RPE does not pass the Painlevé test and hence does not have Painlevé property.

Using the Laurent series we can easily see that, in Eq.(6), the coefficients of the leading term are dominant and thus determined by the left-hand side term $\ddot{R}$ and the first term of the right-hand side $-3\dot{R}^2/2R$ because both of them have the lowest exponent of $\tau$. Eq.(6) is similar to P$_3$ or P$_4$ of the Painlevé transcendents, but complicated by the coefficients. It suggests that the coefficients of a differential equation play a critical role in solving the differential equation analytically. Therefore, the RPE cannot pass the Painlevé test. In the following sections, the possible analytical solutions to RPE type equations are discussed with using symmetry analysis.

**Symmetry Analysis**

Since looking at the domain and codomain of a given differential equation as a product manifold can be naturally generalized to the jet manifold, it provides a more unified and concise language to the Lie symmetry method by studying jet bundles and corresponding Cartan distributions of the differential equation. We plan to discuss the problem here using the machinery of jet bundles. We shall first introduce some basic concepts relevant to the present paper, readers who wish to explore more details can go to (Saunders 1989, Krasil'shchik and Vinogradov 1999)

Let $(E, \pi, M)$ be a locally trivial smooth bundle over a smooth manifold $M$, of which the set of *sections* is $\Gamma(\pi) = \{\sigma : \pi \circ \sigma = id_M\}$. The *r-jet bundle manifold* associated with the bundle can be defined as

$$J_p^r(\pi) = \{j_p^r \sigma : p \in M, \sigma \in \Gamma(\pi)\}$$

where the *r-th* equivalence class $j_p^r \sigma$ is called *r-jet* of $\sigma$ at $p$. Thus, the disjoint union of the fiber manifolds

$$J^r(\pi) = \bigsqcup_{p \in M} j_p^r(\pi)$$

forms a smooth *vector bundle* $(J^r(\pi), \pi_r, M)$ that is called *jet bundle*. The construction also induces *affine bundles* $(J^r(\pi), \pi_{r,k}, J^k(\pi))$, here $0 \leq k < r$ and $J^0(\pi) = E$.

Suppose $\theta \in J^r(\pi)$, Let $\Gamma_\sigma^r \subset J^r(\pi)$ be the graph of r-jets, then the span of all planes tangent to the graph $\Gamma_\sigma^r |_\theta$ is called the *Cartan plane* $\mathcal{C}_\theta$. The disjoint union of the Cartan planes

$$\mathcal{C} = \bigsqcup_{\theta \in J^r(\pi)} \mathcal{C}_\theta$$

is an integrable distribution that is called the *Cartan distribution*, which is the basic geometric structure on the manifold $J^r(\pi)$.

Given an ordinary differential equation as

$$F(t, R, \dot{R}, \ddot{R}) = 0$$

Then the set

$$\mathcal{E} = \{(t, R, \dot{R}, \ddot{R}) : F(t, R, \dot{R}, \ddot{R}) = 0\}$$

defines a submanifold in $J^2(\pi) = (\mathbb{R} \times \mathbb{R}^3, \pi, \mathbb{R})$. The Cartan distribution on the 2-jet manifold in the local coordinates then can be characterized by the following contact forms

$$\omega_0 = dR - \dot{R}dt$$
$$\omega_1 = d\dot{R} - \ddot{R}dt$$

of which the restriction on the equation $\mathcal{E}$ induces the Cartan distribution on the equation (Krasil'shchik and Vinogradov 1999, Xia 2015)

$$\mathcal{C}_\theta(\mathcal{E}) = \mathcal{C}_\theta \cap T_\theta \mathcal{E}$$

where $T_\theta \mathcal{E}$ is the tangent space on the equation at $\theta$. A maximal integral manifold of the Cartan distribution $\mathcal{C}_\theta(\mathcal{E})$ is called a *general solution* to the equation $\mathcal{E}$.

As for the RPE in this paper, we can define a nowhere vanishing vector field on $J^0(\pi)$ spanned by

$$X = \xi(t, R)\frac{\partial}{\partial t} + \eta(t, R)\frac{\partial}{\partial R}$$

which is also generated by a one-parameter group $\Phi \in C^\infty(J^0(\pi))$. Then $X$ is called the *Lie point symmetry* if its lifting on $J^2(\pi)$ is in the form of

$$X^{(2)} = \xi\frac{\partial}{\partial t} + \eta\frac{\partial}{\partial R} + \eta^I \frac{\partial}{\partial \dot{R}} + \eta^{II}\frac{\partial}{\partial \ddot{R}}$$

such that

$$X^{(2)} F(t, R, \dot{R}, \ddot{R})|_{F=0} = 0$$

where

$$\eta^I = D_t\eta - \dot{R}D_t\xi$$
$$\eta^{II} = D_t\eta^I - \ddot{R}D_t\xi$$

and the operators of total differentiation are as

$$D_t = \frac{\partial}{\partial t} + \dot{R}\frac{\partial}{\partial R} + \ddot{R}\frac{\partial}{\partial \dot{R}}$$

Furthermore, if $\Phi \in C^\infty(J^1(\pi))$, we call $X$ the *Lie contact symmetry*. One can quickly check that the lifting of Lie symmetries preserve the Cartan distribution (Krasil'shchik and Vinogradov 1999, Xia 2015), e.g.

$$\mathcal{L}_{X^{(2)}}\omega = \lambda\omega, \qquad \forall \lambda \in C^\infty\left(J^1(M)\right)$$

We know that some differential equations can be solved though they do not admit any lie symmetry. Some authors then proposed new symmetry methods by which their Lie algebras are not closed. In this paper, we remind readers that the so-called *λ-symmetry* (Muriel and Romero 2011) is nothing but with the following expression

$$\eta^I = D_t\eta - \dot{R}D_t\xi + \lambda(\eta - \dot{R}\xi)$$
$$\eta^{II} = D_t\eta^I - \ddot{R}D_t\xi + \lambda(\eta^I - \dot{R}\xi)$$

where $\lambda \in C^\infty(J^1(\pi))$. Since it has not given new results for our equations, we do not present the *λ-symmetry* analysis here.

*Lie point symmetries*

Lie point symmetry analysis of the similar type of the equation has been done by some authors (Maksimov 2004, Tiwari, Pandey et al. 2013), whereas none of the works relates them to the RPE, in the following section, we will present some primary results for the RPE Eq.(4).

Recall the one-parameter Lie transformation group of the form (Xia 2011)

$$\begin{aligned} t^* &= \phi(t, R, \varepsilon) = t + \xi\varepsilon + o(\varepsilon^2) \\ R^* &= \psi(t, R, \varepsilon) = t + \eta\varepsilon + o(\varepsilon^2) \end{aligned} \tag{12}$$

and the corresponding infinitesimal generator (Lie algebra)

$$X = \xi\frac{\partial}{\partial t} + \eta\frac{\partial}{\partial R} \tag{13}$$

where

$$\begin{aligned} \xi &= \xi(t, R) = \frac{d\phi}{d\varepsilon}\bigg|_{\varepsilon=0} \\ \eta &= \eta(t, R) = \frac{d\psi}{d\varepsilon}\bigg|_{\varepsilon=0} \end{aligned} \tag{14}$$

Also, the second prolongation (lifting) of the vector field $X$ can be written as

$$X^{(2)} = \xi \frac{\partial}{\partial t} + \eta \frac{\partial}{\partial R} + \eta^{I} \frac{\partial}{\partial \dot{R}} + \eta^{II} \frac{\partial}{\partial \ddot{R}} \tag{15}$$

By applying

$$X^{(2)} F(t, R, \dot{R}, \ddot{R})|_{F=0} = 0 \tag{16}$$

the determining equations can be obtained. Usually, solving the determining equations is not easier than solving the original differential equations. It is the reason why the Lie group method was not of much practical use before the developing of modern computers. Right now by using CAS software or related packages, we can readily solve the systems of linear partial differential equations. For the RPE Eq.(4), the determining equations obtained by using MAPLE are as follows

$$2\xi_{RR} R - 3\xi_R = 0 \tag{17}$$

$$4\mathrm{Re}\xi_{tR} R^2 - 2\eta_{RR} \mathrm{Re} R^2 - 3\eta_R \mathrm{Re} R - 4\xi_R + 3\eta \mathrm{Re} = 0 \tag{18}$$

$$\mathrm{Re}\xi_{tt} R^{3k+3} - 2\eta_{tR} \mathrm{Re} R^{3k+3} - 3\mathrm{Re}\, p(t) \mathrm{Th}\xi_R R^{3k+2} - 3\eta_t \mathrm{Re} R^{3k+2}$$
$$-3\mathrm{Re}\mathrm{We}\xi_R R^{3k+1} - \xi_t R^{3k+1} + 2\eta R^{3k} + 3\mathrm{p_n}\mathrm{Re}\xi_R R^2 = 0 \tag{19}$$

$$\mathrm{Re}\eta_{tt} R^{3k+3} + 2\mathrm{Re}\mathrm{Th}\, p(t)\xi_t R^{3k+2} + \mathrm{Re}\mathrm{Th}\, \dot{p}(t)\xi R^{3k+2} - \mathrm{Re}\mathrm{Th}\eta_R p(t) R^{3k+2}$$
$$+2\mathrm{Re}\mathrm{We}\xi_t R^{3k+1} - \mathrm{Re}\mathrm{We}\eta_R R^{3k+1} - \mathrm{Re}\mathrm{Th}\eta p(t) R^{3k+1} + \eta_t R^{3k+1} \tag{20}$$
$$-2\mathrm{Re}\mathrm{We}\eta R^{3k} - 2\mathrm{p_n}\mathrm{Re}\xi_t R^2 + \mathrm{p_n}\mathrm{Re}\eta_R R^2 + 3k\mathrm{p_n}\mathrm{Re}\eta R + \mathrm{p_n}\mathrm{Re}\eta R = 0$$

It is not difficult to check in the CAS software MAPLE that, the only solution to above equations is zero with arbitrary $p(t)$ or the equation coefficients. This fact is somewhat the drawback of all kinds of symmetry analysis. Their symmetries are very much sensitive to those coefficients and boundary conditions, and hence it is unable to guarantee the general solutions to a given differential equation. That is the motivation we carry out Lie symmetry analysis on the equations of specific physical meanings. We shall then analyze the possible solutions by specifying coefficients and the unknown function $p(t)$ properly.

**Case 1:** $p(t) = \mathrm{const} = p_0$. In this case, the physical meaning corresponds to the behaviors of a gas bubble under hydrostatic pressure. Then a solution to the above four determining equations is

$$\xi = \mathrm{const}, \quad \eta = 0 \tag{21}$$

It correlates to a one-dimensional Lie subalgebra

$$X = \text{const}\frac{\partial}{\partial t} \tag{22}$$

which is corresponding to a translation group along *x*-axis. An invariant solution for the equation thus can be obtained

$$R = \frac{1}{3p_0 \text{Th}}\left(\frac{Y}{2} + \frac{2\text{We}^2}{Y} - \text{We}\right)$$

and (23)

$$Y = \left(43^{\frac{3}{2}} p_0 \text{Th}\sqrt{p_n\left(27 p_0^2 p_n \text{Th}^2 - 4\text{We}^3\right)} - 8\text{We}^3 + 108 p_0^2 p_n \text{Th}^2\right)^{\frac{1}{3}}$$

It is the equilibrium radius of the bubble at hydrostatic pressure $p_0$. Furthermore, the Lie subalgebra allows reduction of the equation into the following first order ODE

$$2x^2 y \frac{dy}{dx} + 3xy^2 + \frac{2}{\text{Re}}y - 2p_n x^{1-3k} + 2p_0 \text{Th} x + 2\text{We} = 0 \tag{24}$$

here we denote

$$y(x) = \dot{R}(t), \quad x = R(t) \tag{25}$$

as a matter of convenience.

*Case 1.1:* The viscosity and surface tension are not included. Thus, Eq.(24) is reduced to

$$2xy\frac{dy}{dx} + 3y^2 - \frac{2p_n}{x^{3k}} + 2p_0 \text{Th} = 0 \tag{26}$$

Let the initial condition to be $\dot{R}(0) = 0$ or $y(1) = 0$, we will have

$$y(x)^2 = \frac{2}{3}p_0 \text{Th}\left(\frac{1}{x^3} - 1\right) + \frac{2}{3}\left(\frac{1}{x^3} - \frac{1}{x^{3k}}\right)\frac{p_n}{(k-1)} \tag{27}$$

or

$$\dot{R}^2 = \frac{2}{3}p_0 \text{Th}\left(\frac{1}{R^3} - 1\right) + \frac{2}{3}\left(\frac{1}{R^3} - \frac{1}{R^{3k}}\right)\frac{p_n}{(k-1)} \tag{28}$$

When the polytropic constant $k=1$, the above solution (28) is reduced to the previous result (Mancas and Rosu 2016). Substituting the above equation into Eq.(4), and applying an additional initial condition $R(0)=1$, we have

$$t = \sqrt{\frac{3}{2}} \int_R^1 \frac{a^{3/2}}{\sqrt{\left((1-a^3)\text{Th}\, p_0 - p_n \frac{(a^{3-3k}-1)}{(k-1)}\right)}} da \qquad (29)$$

Eq.(29) characterizes how the radius of a bubble evolved under varying hydrostatic pressures. It is a more general case considering the impact of polytropic constant, which could be substantial in engineering applications. For example, a gas bubble in a liquid exchanges gas with the surrounding liquid and thus the gas content inside the bubble also alters. And Eq.(29) is capable of characterizing how the evolution of the radius is affected by the gas content.

*Case 1.2:* Continuing on the Rayleigh equation, if the excitation pressure has the form of

$$p(t) = c(at+b)^{-\frac{6k}{3k+2}} \qquad (30)$$

which corresponds to the acoustic field inside an experimental flask (Maksimov 2004). In this situation, its determining equations can be obtained as

$$\xi(t,R) = at+b, \quad \eta(t,R) = \frac{2aR}{3k+2} \qquad (31)$$

where $a$, $b$ and $c$ are arbitrary constants. For a special case $a=b=c=k=1$, we have $p(t) = (t+1)^{-6/5}$. The invariant solution to this case is

$$R(t) = \left(\frac{p_n}{\text{Th}}\right)^{\frac{1}{3}} (t+1)^{\frac{2}{5}} \qquad (32)$$

**Case 2:** $k=2/3$, $We=0$. In this case, we manually set the polytropic constant $k<1$, which may not refer to any explicit physical meaning. However, it is the case that the solvable Lie subalgebra can be obtained. Meanwhile, Given the excitation pressure

$$p(t) = \frac{c}{(at+b)} \qquad (33)$$

then we have

$$\xi(t,R) = at+b, \quad \eta(t,R) = \frac{a}{2}R \tag{34}$$

and the Lie subalgebra is

$$X = (at+b)\frac{\partial}{\partial t} + \frac{a}{2}R\frac{\partial}{\partial R} \tag{35}$$

For the special case $a = b = c = 1$, we have the invariant solution in the form of

$$R(t)^2 = 2\left(\sqrt{\left(2\text{Th}+\frac{1}{\text{Re}}\right)^2 + 2p_n} - 2\text{Th} - \frac{1}{\text{Re}}\right)(t+1) \tag{36}$$

**Case 3:** $k = 1/3$, $1/\text{Re} = 0$. In this case, we neglect the impact of viscosity and assume the excitation pressure is in the form of

$$p(t) = \frac{c}{(at+b)^{\frac{2}{3}}} \tag{37}$$

we have

$$\xi(t,R) = at+b, \quad \eta(t,R) = \frac{2a}{3}R \tag{38}$$

and the Lie subalgebra is

$$X = (at+b)\frac{\partial}{\partial t} + \frac{2a}{3}R\frac{\partial}{\partial R} \tag{39}$$

again let $a = b = c = 1$, the invariant solution to this case is

$$R(t) = \frac{1}{2}\left(Y - \frac{3\text{Th}}{Y}\right)(t+1)^{\frac{2}{3}}$$

and  (40)

$$Y = \left(3\sqrt{9(\text{We}-p_n)^2 + 3\text{Th}^3} - 9\text{We} + 9p_n\right)^{\frac{1}{3}}$$

**Discussion**

We have shown that, from the abovementioned analysis, the RPE does not pass the Painlevé test nor admit Lie symmetry groups generally. However, the particular cases of the Rayleigh-Plesset equation have invariant solutions corresponding to several physical meanings. We may also analyze the integrability of the equation by reviewing it from the physical perspective. Multiplying Eq.(4) by the integral factor $R^2\dot{R}$ and assuming $k=1$, we will have

$$\frac{1}{2}\frac{d}{dt}\left(R^3\dot{R}^2\right)+\frac{\dot{R}}{\text{Re}}\frac{1}{2}\frac{d}{dt}\left(R^2\right)+\text{We}\frac{1}{2}\frac{d}{dt}\left(R^2\right)+\text{Th}p(t)R\frac{1}{2}\frac{d}{dt}\left(R^2\right)-\text{p}_\text{n}\frac{d}{dt}\left(\ln R\right)=0 \quad (41)$$

The above form of the Rayleigh-Plesset equation is nothing but the balance of energy. The first term refers to the kinetic energy of a bubble; the second term relates to the interface dissipation due to the viscosity of the surrounding medium; the third term refers to potential energy of the surface tension; the forth term refers to the input energy; and the last term refers to the potential energy of the gas inside the bubble. Rearrange the order of the above equation

$$\frac{1}{2}\frac{d}{dt}\left(R^3\dot{R}^2\right)+\left(\frac{\dot{R}}{\text{Re}}+\text{Th}p(t)R\right)\frac{1}{2}\frac{d}{dt}\left(R^2\right)+\text{We}\frac{1}{2}\frac{d}{dt}\left(R^2\right)-\text{p}_\text{n}\frac{d}{dt}\left(\ln R\right)=0 \quad (42)$$

Therefore, in Eq.(42) we can easily see that the coefficient of the second term $\dot{R}/\text{Re}+\text{Th}p(t)R$, which also depends on the time variable and cannot be written into the time derivative, determines possible first integrations of the Rayleigh-Plesset equation.

**Conclusions**

The integrability of the Rayleigh-Plesset equation was studied using both the Painlevé test and Lie symmetry analysis. Although the equation generally cannot be transformed into any of the six Painlevé equations, solutions to some simplified Rayleigh-Plesset equation are still achievable. Several invariant solutions of specific physical meanings were presented. The physical parameters (as the coefficients in the equation) have a significant impact on the possibility of obtaining analytical solutions. The Rayleigh-Plesset equation does not admit any Lie symmetry groups when the value of the polytropic constant is higher than one; meanwhile, the undetermined function $p(t)$ contributes to the complexity of Lie symmetry analysis. Thus, the invariant solutions can be only obtained by choosing proper coefficients and the unknown function $p(t)$.